\documentclass[journal]{IEEEtran}
\usepackage{graphicx}
\usepackage[normalem]{ulem}
\usepackage{amssymb}
\usepackage{amsmath}
\usepackage{amstext}
\usepackage{amsfonts}
\usepackage[normalem]{ulem}
\usepackage[brazil]{babel}
\usepackage{epstopdf}
\usepackage{amsmath}
\usepackage{subcaption}
\usepackage{amssymb}
\usepackage{url}
\usepackage{here}
\usepackage{hyperref}
\newcommand*{\doi}[1]{\href{http://dx.doi.org/#1}{doi: #1}}
\usepackage{mathtools}
\usepackage{arydshln}
\usepackage{pstricks}

\newtheorem{definition}{Definition}

\renewcommand\thetable{\thesection.\arabic{table}}
\newcommand{\defeq}{\vcentcolon=}

\begin{document}
\renewcommand*\abstractname{Abstract}
\renewcommand{\figurename}{Figure}
\renewcommand{\tablename}{Table}
\renewcommand{\refname}{References}

\title{Privatization of Probability Distributions by the Wavelet Integral approach
 }

\author{Helio M. de Oliveira$^1$, Raydonal Ospina$^1$, Victor Leiva$^2$, Carlos Martin-Barreiro$^3$, Christophe Chesneau$^4$\\ 
$^1$Department of Statistics, CAST Laboratory, Universidade Federal de Pernambuco, CCEN-UFPE, Recife, Brazil\\
$^2$School of Industrial Engineering, Pontificia Universidad Cat\'{o}lica de Valpara\'{i}so, Chile\\
$^3$Faculty of Natural Sciences and Mathematics, Guayaquil, Ecuador\\
$^4$Department of Mathematics, Université de Caen Basse-Normandie, Caen, France}

\date{2015}

\maketitle

\begin{abstract}

\textbf{A naive theory of additive perturbations on a continuous probability distribution is presented. We propose a new privatization mechanism based on a naive theory of a perturbation on a probability using wavelets, such as a noise perturbs the signal of a digital image sensor. The cumulative wavelet integral function is defined and {builds} up the perturbations with the help of this function. We show that an arbitrary distribution function additively perturbed is still a distribution function, which can be seen as a privatized distribution, with the privatization mechanism being a wavelet function. It is shown that an arbitrary cumulative distribution function added to such an additive perturbation is still a cumulative distribution function. Thus, we offer a mathematical method for choosing a suitable probability distribution to data by starting from some guessed initial distribution. The areas of artificial intelligence and machine learning are constantly in need of data fitting techniques, closely related to sensors. The proposed privatization mechanism is therefore a contribution to increasing the scope of existing techniques.} 
\end{abstract}

\providecommand{\keywords}[1]{\textbf{\textit{Index terms---}}}

\section{Introduction}
Given an arbitrary random variable with a continuous cumulative probability distribution (CDF) $F_X(x)$, 
let us consider an additive perturbation $\varepsilon(x)$ so that 
\begin{equation}\label{eq:perturbation}
F_{new}(x):=F_X(x)+\varepsilon(x).
\end{equation}
However, the choice of the disturbance cannot be arbitrary because it could lead to breaking the requirements to deal only with a probability distribution. The following conditions must be hold by the perturbation, namely:
\begin{subequations}
	\begin{equation}
	\lim_{|x| \to \infty }\varepsilon(x)=0.
	\end{equation}
	\begin{equation}
	\varepsilon(x)\leq \left | F_X(x) \right |.
	\end{equation}
	\begin{equation}
	\int_{-\infty }^{\infty }\varepsilon(x) {\rm d}x=0.
	\end{equation}
\end{subequations}
In order to propose a manageable perturbation, let us deal just with compactly supported wavelets, ${\rm supp}~\psi (x)\equiv[a,b]$.\\
\begin{definition}\label{def:cumulative}
(wavelet cumulative function) The wavelet cumulative function $\Psi(x)$ is defined by
	\begin{equation}
	\Psi (x) := \int_{-\infty }^{x}\psi (\zeta ) {\rm d}\zeta.
	\end{equation}
	
Since only continuous compactly supported wavelets are considered, this can be simplified to
\begin{equation}
\Psi (x) = \int_{a}^{x}\psi (\zeta ) {\rm d}\zeta, ~~x \geq a,
\end{equation}
ant the following properties can easily been verified:
\begin{subequations} \label{eq:boundary}
	\begin{equation}
	\Psi (-\infty) =\Psi (x) = \Psi (a)=0, ~~~x\leq a,
	\end{equation}
	\begin{equation}
	\Psi (x) =\Psi (b) = \Psi (\infty) = 0, ~~~~x \geq b. 
	\end{equation}
	\begin{equation}
	\frac{\mathrm{d} \Psi (x)}{\mathrm{d} x}=\psi (x).
	\end{equation}
\end{subequations}
To begin with, let us deal with the distribution $\mathcal{U}[0,1]$, which CDF is given by 
\begin{equation}
F_X(x)=x,~~0\leq x \leq 1.
\end{equation}
A  (afin) mapping is proposed to bring the support [0,1] of the uniform distribution with the support $[a,b]$ of the wavelet: $[0,1]\overset{\rm map}{\rightarrow}[a,b]$. Then we propose to chose a particular perturbation $\varepsilon(x)$ according to
\begin{equation}
\varepsilon(x) =\Psi_{[0,1]}(x):=\frac{1}{b-a}\Psi \left ( (b-a)x+a \right ).
\label{eq:perturbation_def}
\end{equation}
For this particular choice, the new distribution defined in Eqn~\ref{eq:perturbation} has the same support as the original distribution with no perturbation added. Furthermore, imposing the condition:
\begin{equation}
|\psi (t)|\leq 1,
\end{equation}
it follows that
\begin{equation}
\left | \varepsilon(x) \right | \leq \frac{1}{b-a}\int_{a}^{(b-a)x+a} |\psi (\zeta )| {\rm d}\zeta,
\end{equation}
 so that we guarantee that for $x \in [0,1]$
\begin{equation}
\left | \varepsilon(x) \right |\leq x.
\end{equation}
Therefore, the condition $F_{new}(x)\geq 0$ is assured. Then, we need to see whether the $F_{\rm new}(x)$ is always a non-descending function or not. So we examine the behavior of the corresponding probability density function (pdf) given by 
\begin{equation}
f_{new}(x)=\frac{\mathrm{d} F_{new}}{\mathrm{d} x}=1+\frac{1}{b-a}.\frac{\mathrm{d} \Psi \left ( (b-a)x+a \right )}{\mathrm{d} x},
\end{equation}
implying
\begin{equation}
f_{new}(x)=1+ \psi \left ( (b-a)x+a \right ).
\end{equation}
It follows that $f_{new}(x)\geq 0$, and that $\int_{-\infty }^{\infty }f_{new}(\zeta ){\rm d}\zeta =1$, thereby proving that this is indeed a valid PDF to be considered.
\end{definition}

\begin{figure}
	\centering
	\begin{subfigure}{0.4\textwidth}
		\includegraphics[scale=0.5]{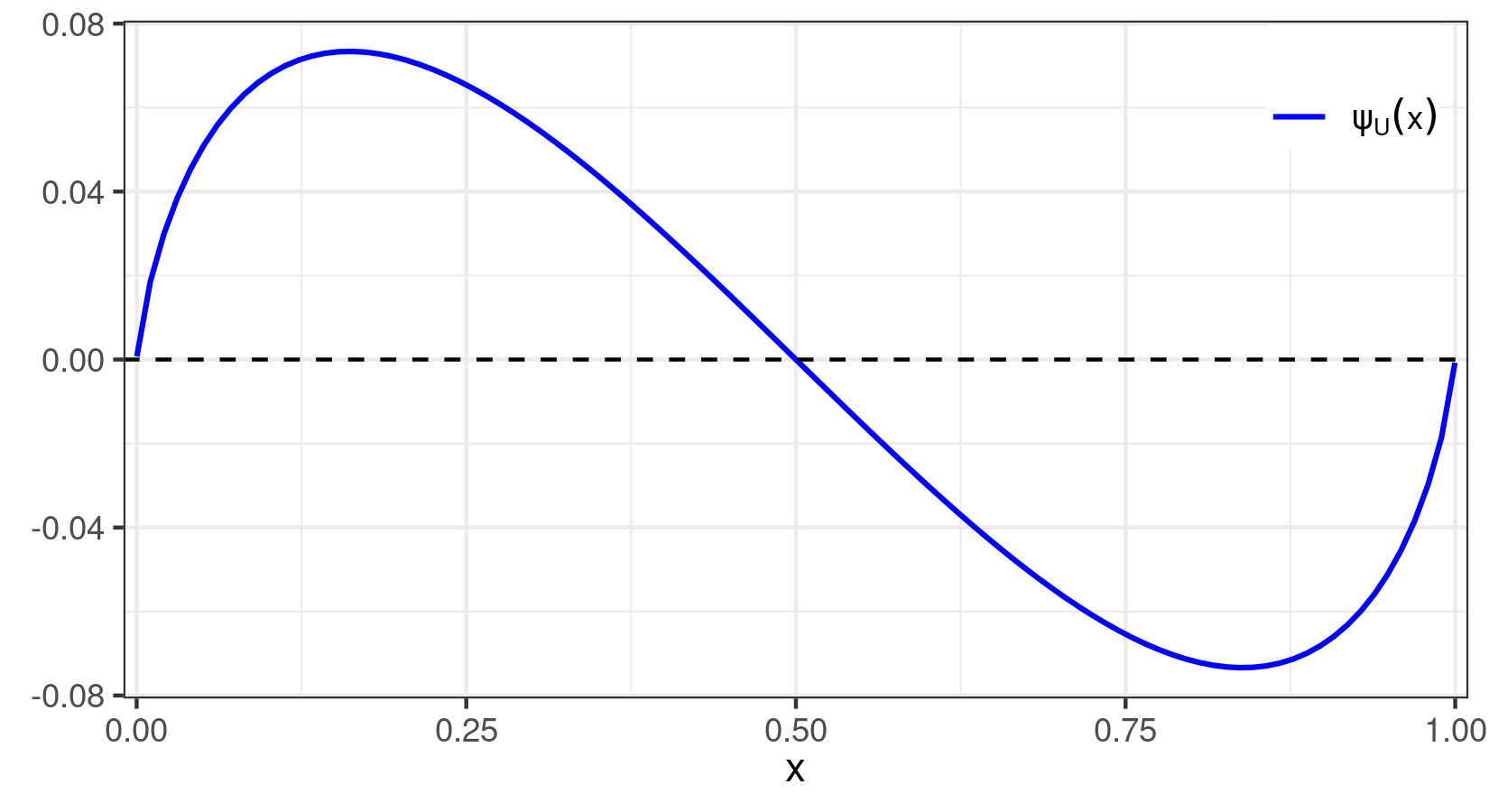}
		\captionof{figure}{Wavelet $\psi_U(x)$  for modeling a perturbation associated with the uniform distribution.}
	\end{subfigure}%
	\hfill
	\begin{subfigure}{0.4\textwidth}
		\includegraphics[scale=0.5]{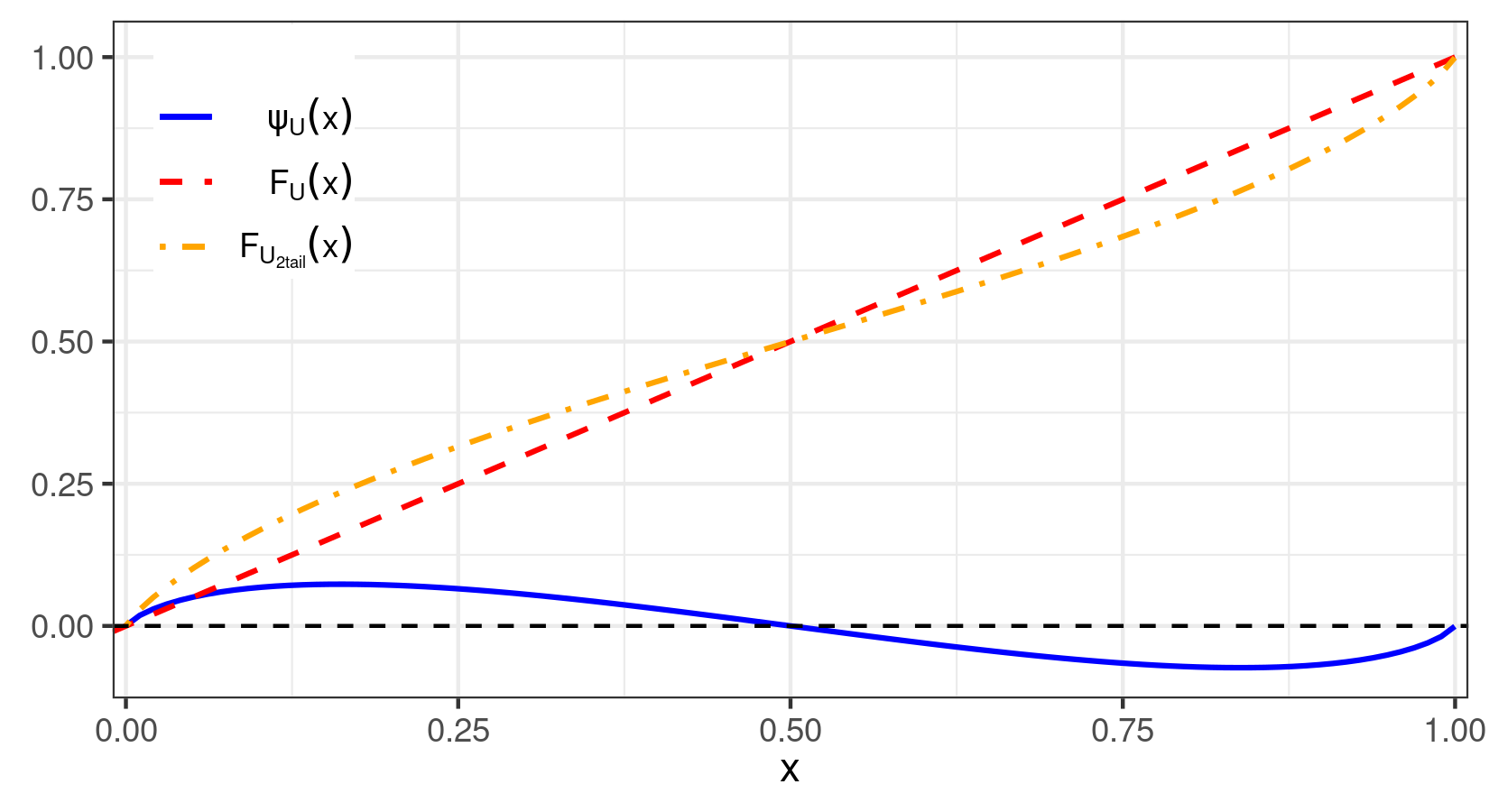}
		\captionof{figure}{Perturbed uniform distribution (\textit{dashed line}), and uniform distribution (\textit{dotted line}).}
	\end{subfigure}
	\caption{Perturbed uniform distribution (dotted line), and uniform distribution (dashed line).}
	\label{fig:fig_waveletpsiU}
\end{figure}

\begin{figure}[!ht]
	\centering
	\begin{subfigure}{0.4\textwidth}
		\includegraphics[scale=0.5]{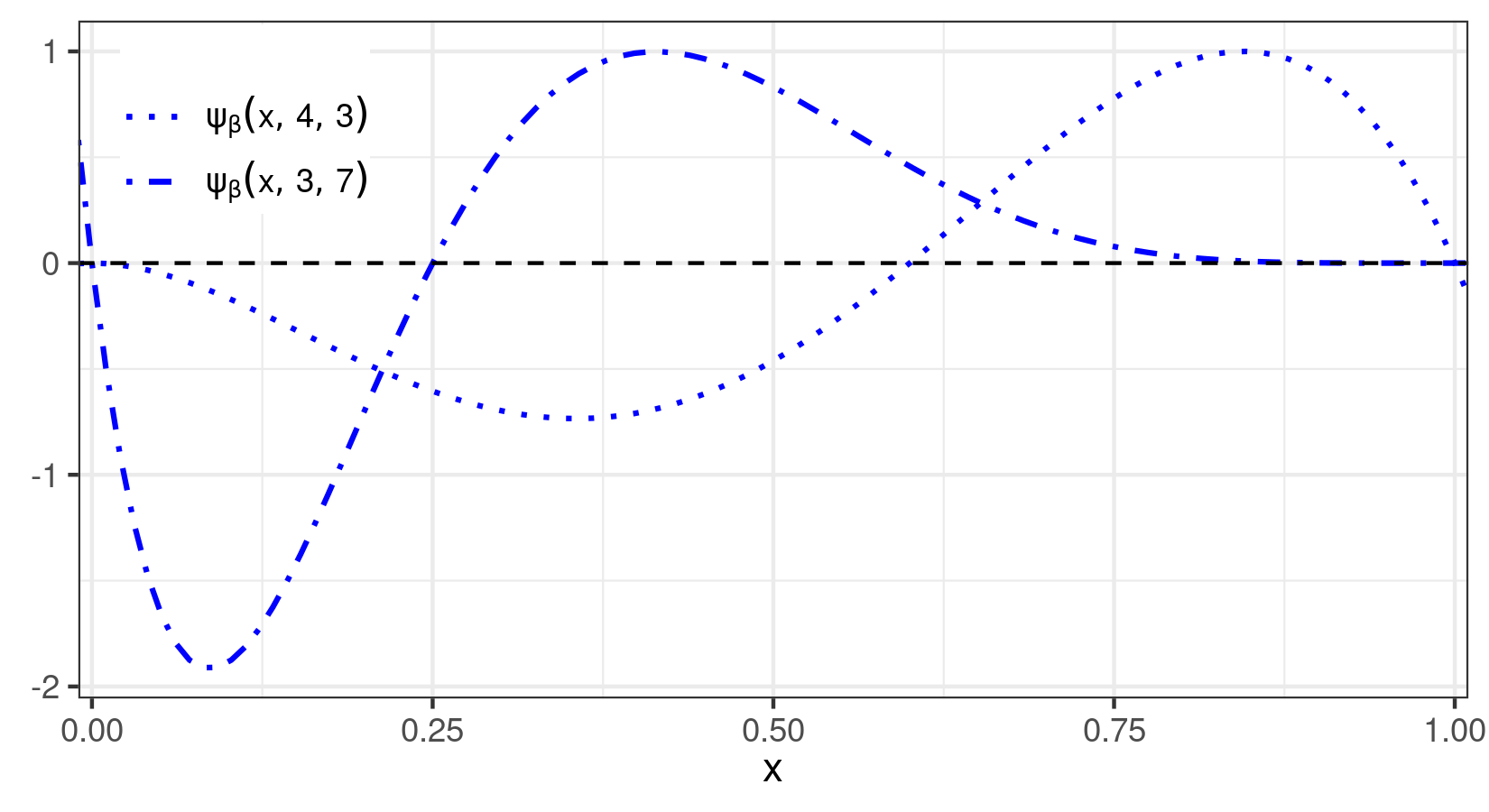}
		\captionof{figure}{Beta-wavelet perturbation associated with the uniform distribution $\mathcal{U}[0,1],$ $\psi_{beta}(x,4,3)$ (dotted)  and $\psi_{beta}(x,3,7)$ (dotted line). }
	\end{subfigure}%
	\hfill
	\begin{subfigure}{0.4\textwidth}
		\includegraphics[scale=0.5]{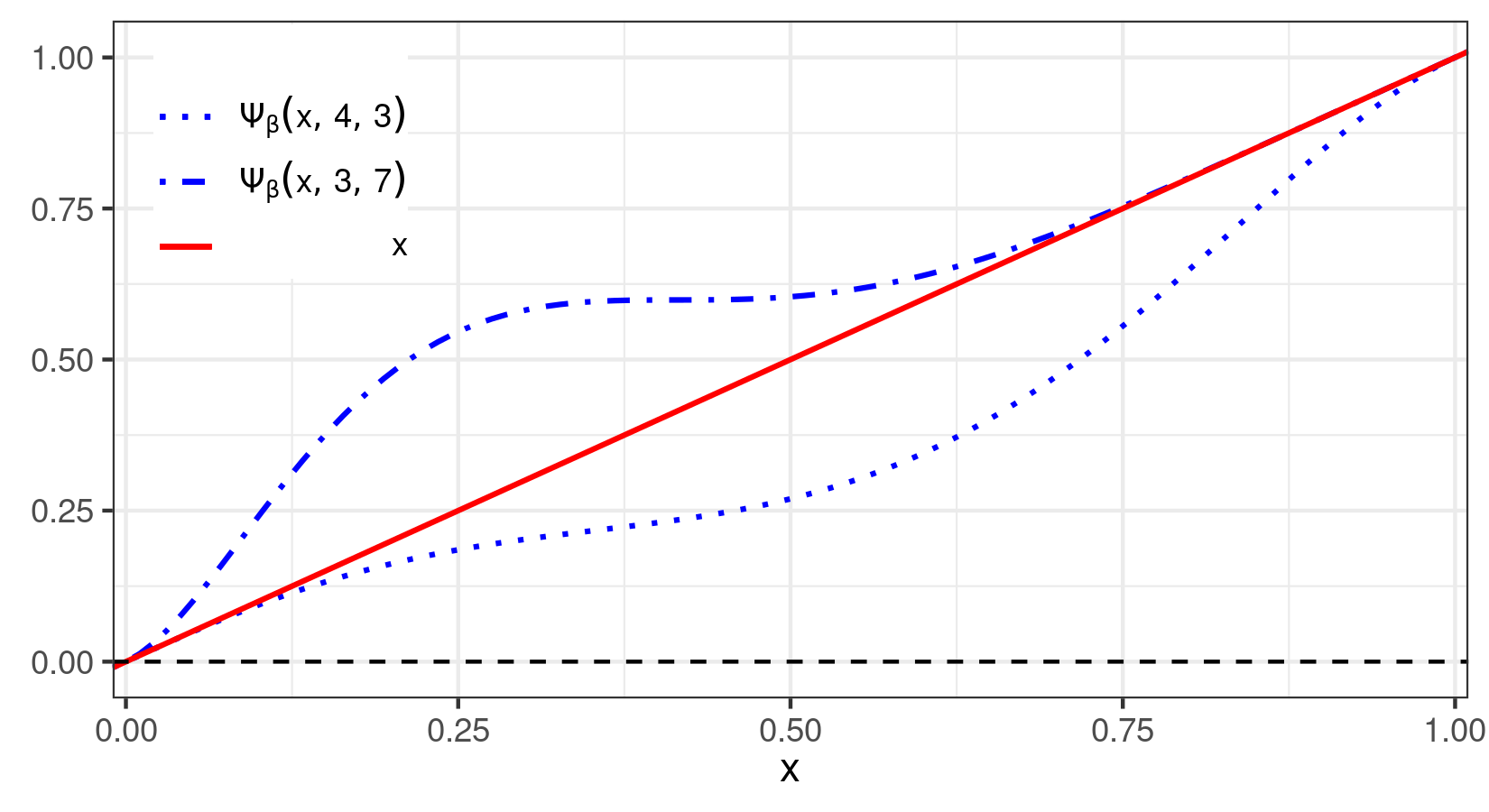}
		
		\captionof{figure}{ Perturbed distribution $\psi_{beta}(x,4,3)$ (dotted), $\psi_{beta}(x,3,7)$ (dotted line)  and original uniform distribution (line).}
	\end{subfigure}
	\caption{Perturbations $\psi_{\rm beta}(x,3,4)$ and $\psi_{\rm beta}(x,3,7)$ of a uniform CDF.}
	\label{fig:fig_waveletbeta}
\end{figure}

Let us first assume a compactly supported wavelet $\psi_U(x)$ defined within $[0,1]$ proposed in \cite{de_Oliveira} formulated as
\begin{equation} \label{eq:psiU}
\psi_U(x) \defeq -\frac{1}{2}x\ln(x)+\frac{1}{2}(1-x)\ln(1-x),
\end{equation}
Fig.~\ref{fig:fig_waveletpsiU} shows the original ($\mathcal{U}$[0,1]-uniform) distribution and the new one generate by the perturbation identified in Eqn~\ref{eq:psiU}.\\
Another compactly supported wavelets family with parameters that can be adjusted is the beta wavelet family 
\cite{de_Oliveira-Araujo}. 
One of the advantages of adopting beta wavelet perturbations consists of the {easy replacement} of parameters $\alpha$ and $\beta$ to shape the perturbation $\psi_{\rm beta}(x,\alpha,\beta)$.  {Figure} \ref{fig:fig_waveletbeta} shows the local perturbation generated by beta wavelets for a few selected parameters. 
The advantage of taking this wavelet family is the simple parametrization {that} drives the asymmetry of the resulting probability distribution. The parametric plot between the uniform and beta wavelet perturbed CDFs for the two {cases} is shown in {Figure} \ref{fig:fig_waveletbeta}. 
This approach can be employed to introduce asymmetries in a chosen probability distribution, controlling through the beta wavelet parameter. 

\begin{figure}[!ht]
	\centering
	\begin{subfigure}{0.4\textwidth}
		\includegraphics[scale=0.5]{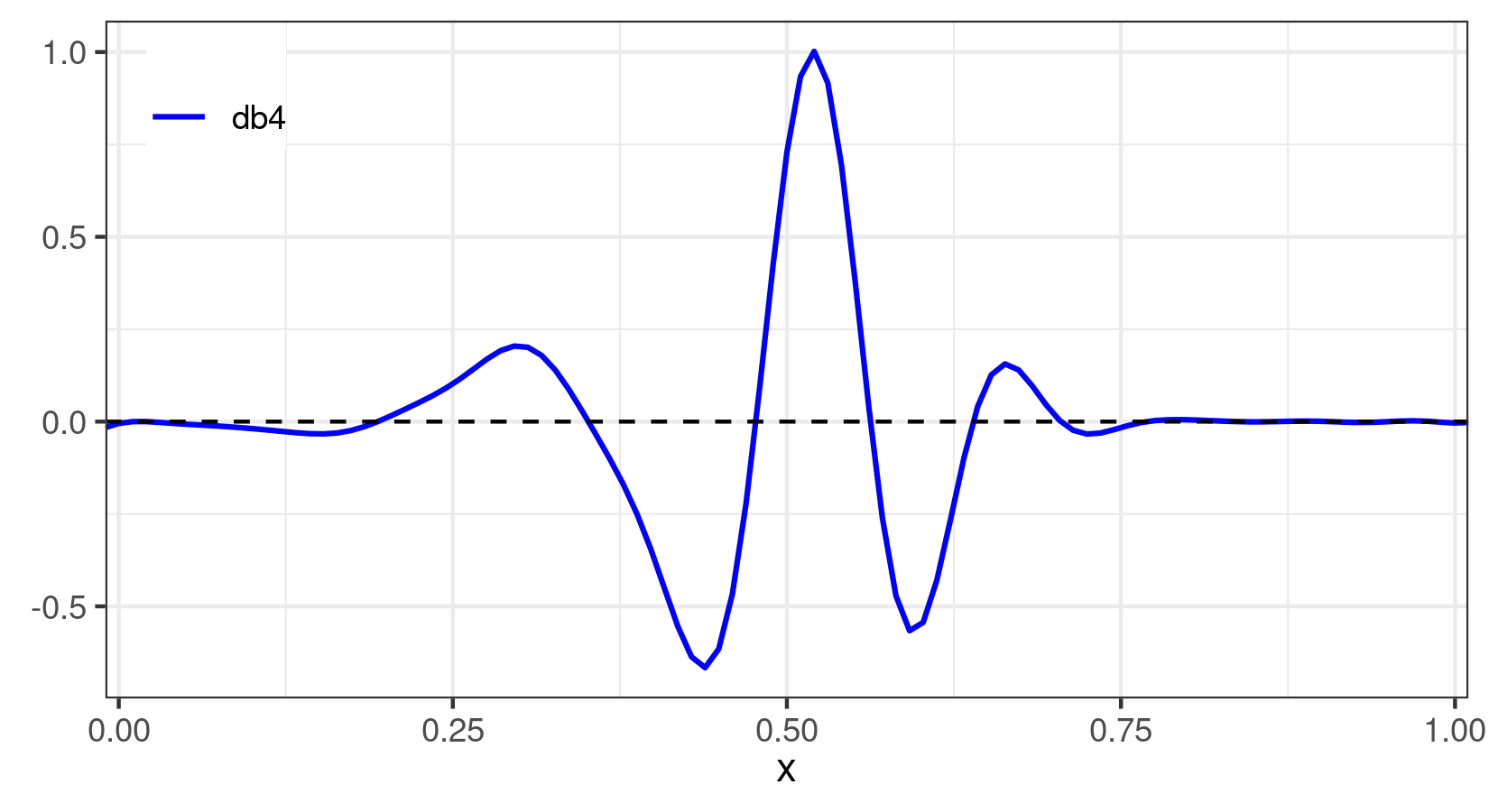}
		\captionof{figure}{Daubechies-wavelet engendering  $\varepsilon(x)$-perturbation associated with the uniform distribution: db4.}
	\end{subfigure}%
	\hfill
	\begin{subfigure}{0.4\textwidth}
		\includegraphics[scale=0.5]{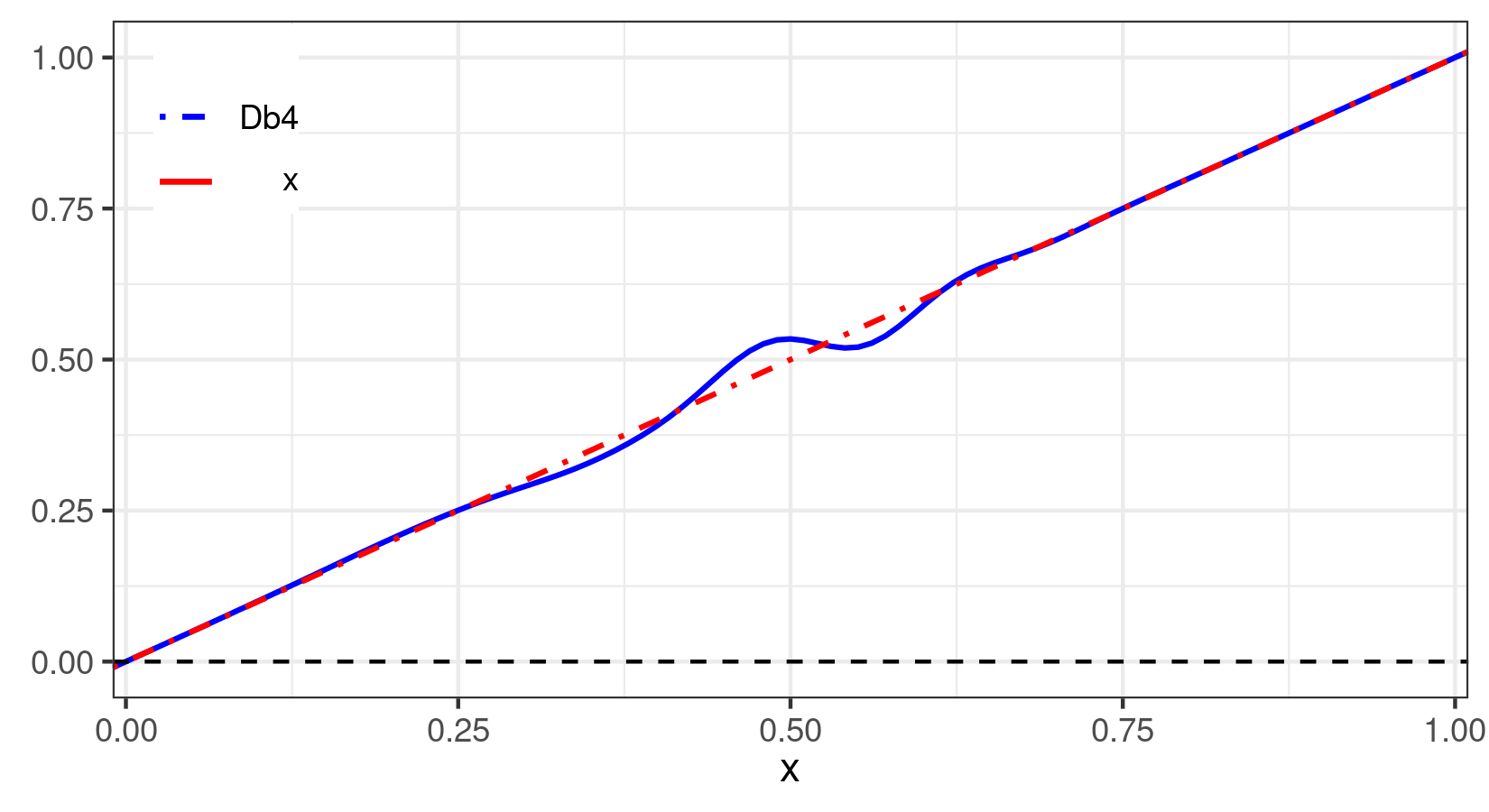}
		\captionof{figure}{Perturbed distribution (\textit{line}), and uniform distribution (\textit{dotted line}) (Matlab generated).}
	\end{subfigure}
	\caption{$db4$-perturbation on a Uniform Probability Distribution Function $\mathcal{U}[0,1]$.}
	\label{fig:fig_waveletdaub}
\end{figure}

Among the compactly supported wavelets, certainly the most used are the wavelets of Daubechies. 
Expressions {close} to approximate Daubechies wavelets of any order have been proposed in \cite{Vermehren-deO}. Here these continuous approximation were used to plot the db4 perturbation adapted to a uniform distribution [0,1], using commands in Matlab${}^{\tt TM}$ (see  Fig.~\ref{fig:fig_waveletdaub}a).

\begin{figure}[!ht]
	\centering
	\begin{subfigure}{0.4\textwidth}
		\includegraphics[scale=0.5]{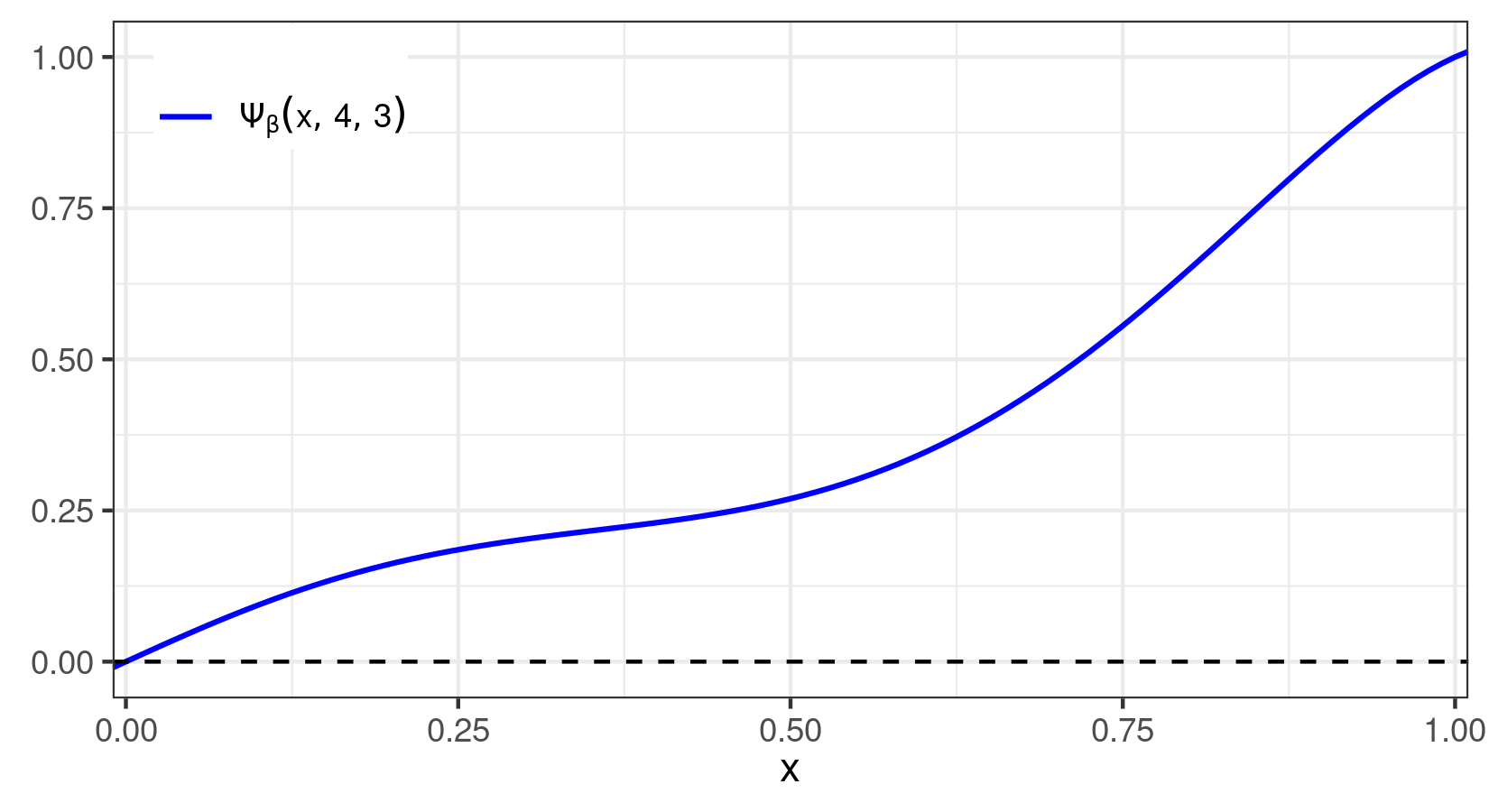}
		\captionof{figure}{Parametric plot between the two CDF: uniform and $\psi_{\rm beta}(x,4,3)$ perturbation.}
	\end{subfigure}%
	\hfill
	\begin{subfigure}{0.4\textwidth}
		\includegraphics[scale=0.5]{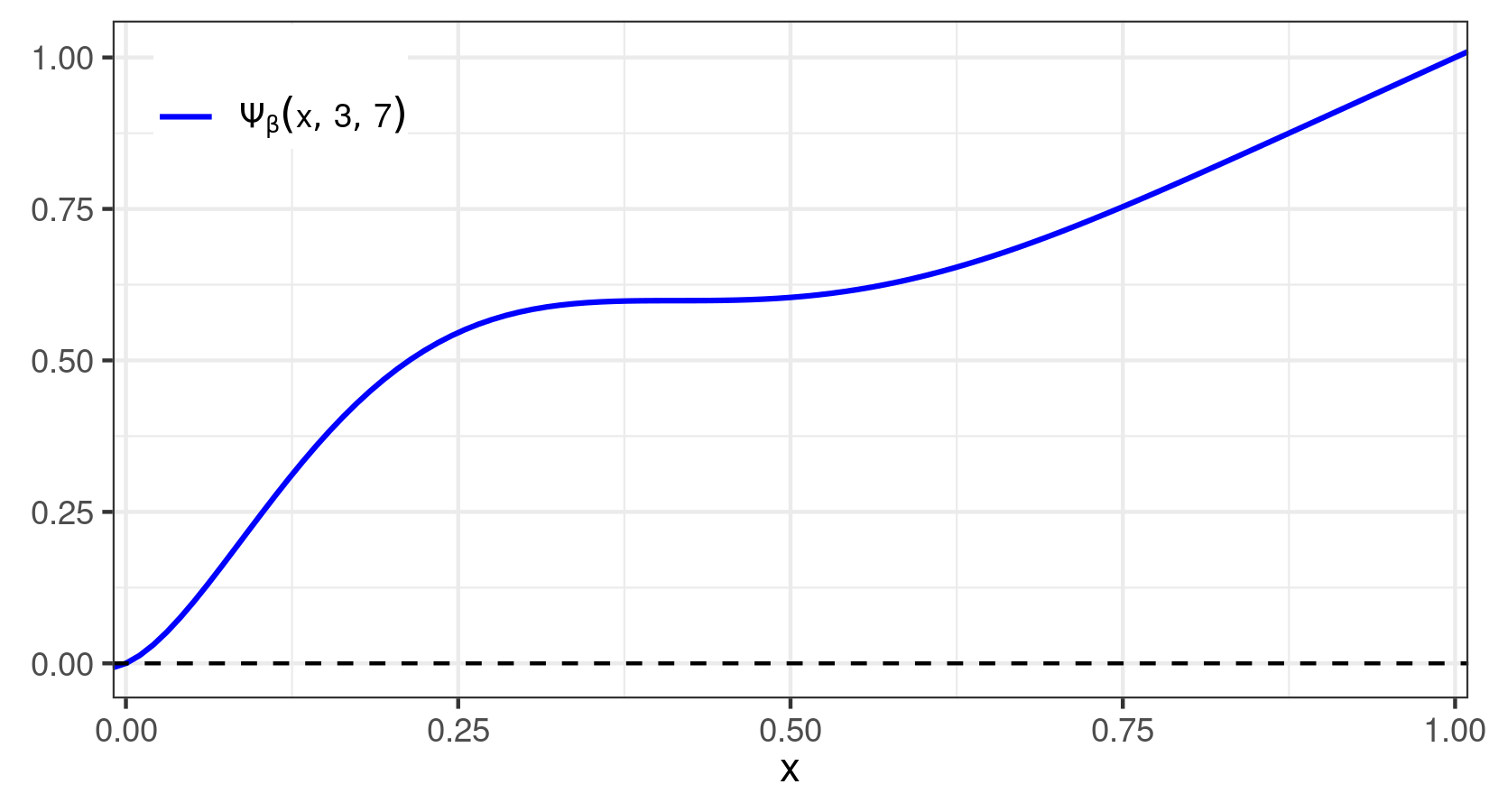}
		\captionof{figure}{Parametric plot between the two CDF: uniform and  $\psi_{\rm beta}(x,3,7)$ perturbation.}
	\end{subfigure}
	\caption{Perturbation $\psi_{\rm beta}(x)$ of a uniform probability distribution function.}
	\label{fig:fig_parametric}
\end{figure}

\begin{figure}[!ht]
	\centering
	\begin{subfigure}{0.4\textwidth}
		\includegraphics[scale=0.5]{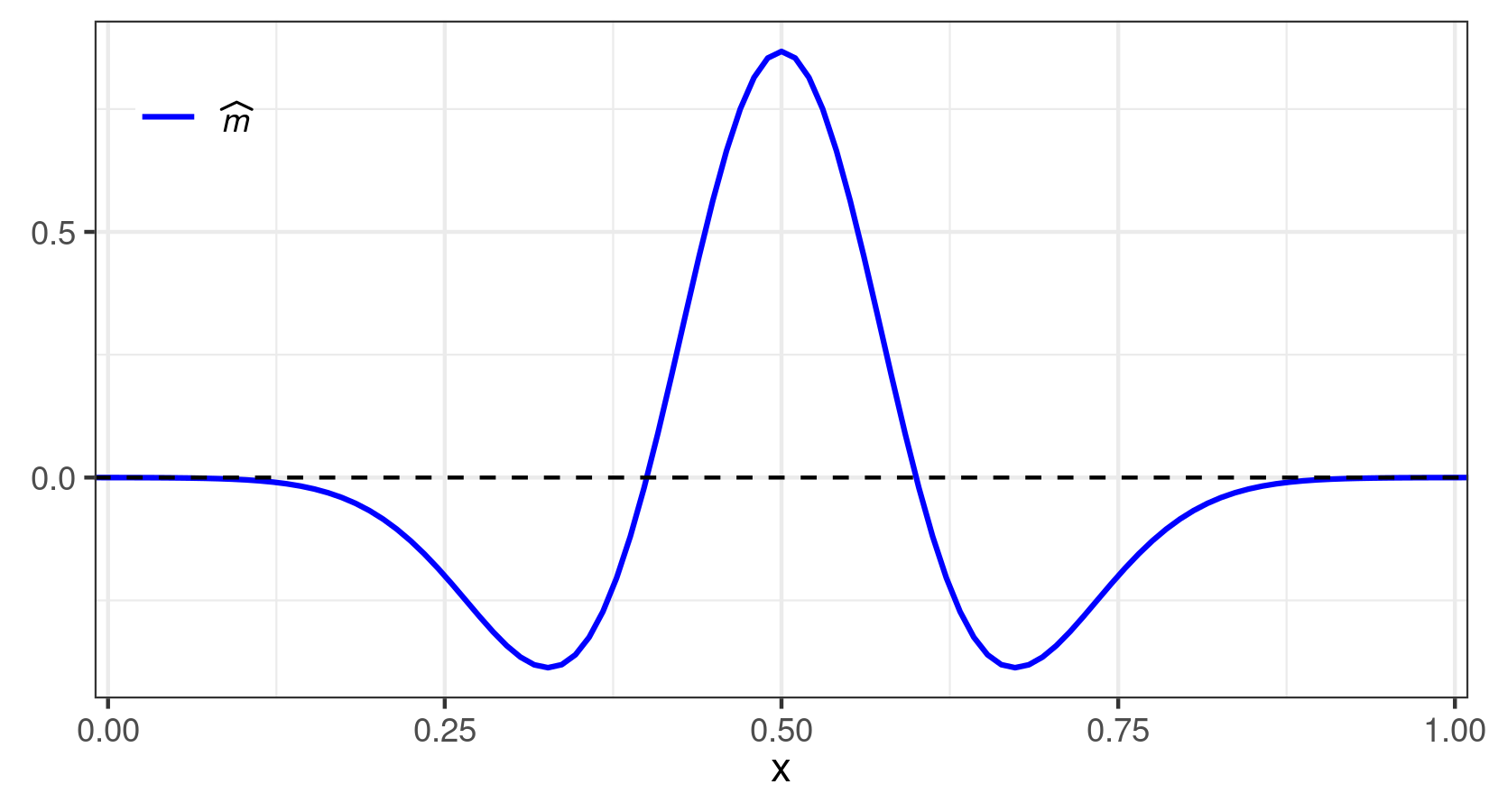}
		\captionof{figure}{Sombrero-wavelet engendering  $\varepsilon(x)$-perturbation associated with the uniform distribution: mex-hat.}
	\end{subfigure}%
	\hfill
	\begin{subfigure}{0.4\textwidth}
		\includegraphics[scale=0.5]{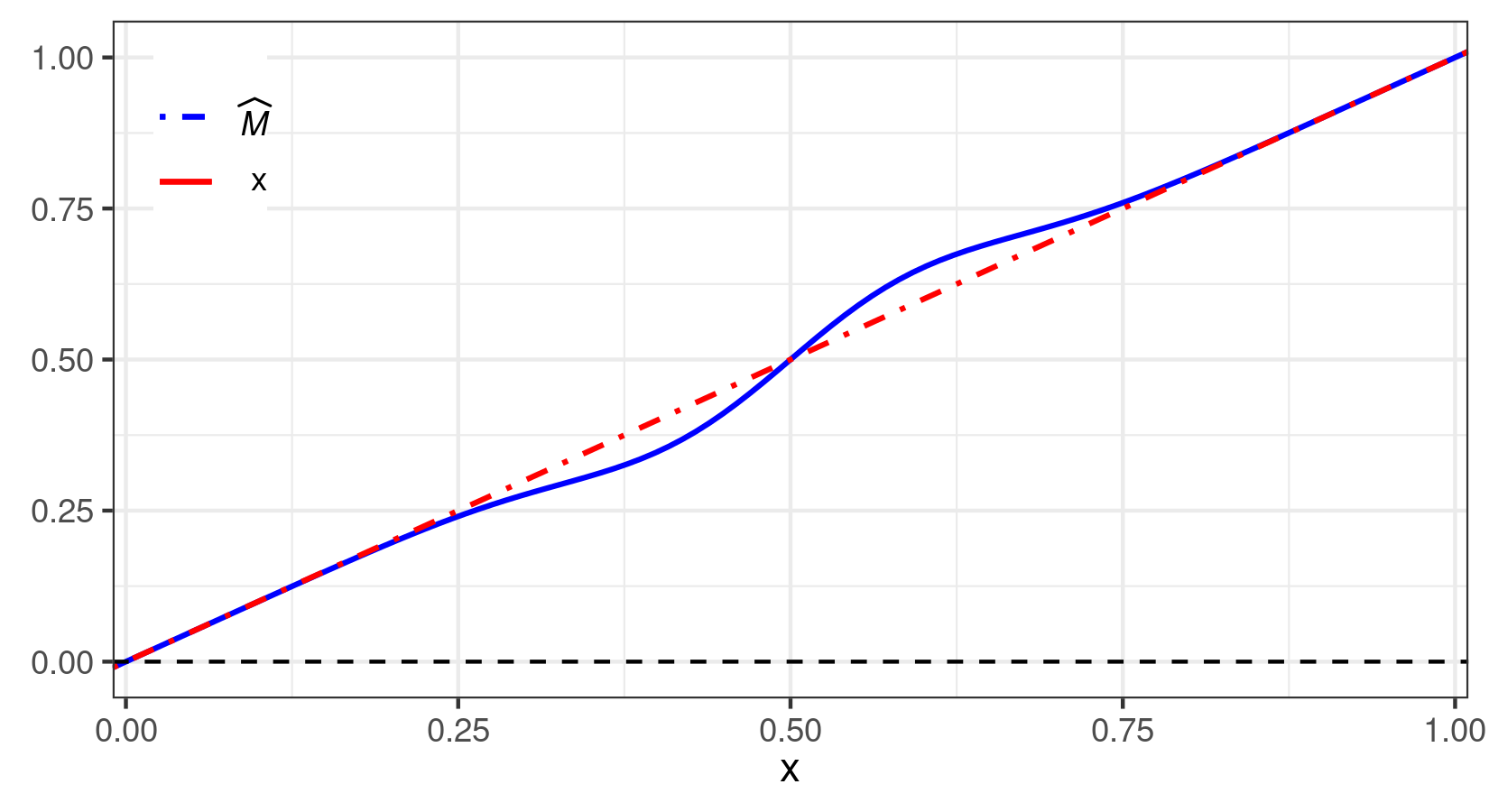}
		\captionof{figure}{Mexican-hat-perturbed distribution (\textit{line}), and uniform distribution (\textit{dashed line}).}
	\end{subfigure}
	\caption{Sombrero-perturbation on a Uniform Probability Distribution Function $\mathcal{U}[0,1]$.}
	\label{fig:fig_waveletmexhat}
\end{figure}
\section{Choosing a Perturbation to an Arbitrary Probability Distribution}
Now, we offer a valid perturbation for an arbitrary CDF $F_X$. For a given compactly supported wavelet $\psi$ with a  wavelet cumulative function (see Definition \ref{def:cumulative}), consider a new chosen distribution according to
\begin{equation} 
F_{new}(x)=F_X(x)+\frac{1}{b-a}\frac{\Psi \left ( (b-a).F_X(x)+a \right )}{\max_{\zeta \in {\rm supp}\psi }|\psi (\zeta)|}.
\end{equation}
It is promptly seen that $F_{new}(-\infty)=0$ and $F_{new}(\infty)=1$, which is a consequence of 
\begin{equation}
f_{new}(x)=f_X(x)+\frac{\psi \left ( (b-a) F_X(x)+a \right )}{\max_{\zeta \in {\rm supp}\psi }|\psi (\zeta)|} f_X(x).
\end{equation}
This equation results in $0 \leq \left | \varepsilon (x) \right | \leq f_X(x)$ so this is actually a valid perturbation. Clearly,
\begin{equation}
\int_{-\infty }^{\infty }f_{new}(\zeta ){\rm d}\zeta =\int_{-\infty }^{\infty }f_X(\zeta ){\rm d}\zeta=1,
\end{equation}\\
since that
\begin{equation}
\frac {1}{\max_{\zeta \in supp\psi } |\psi(\zeta)|} \int_{-\infty }^{\infty } \psi \left ( (b-a)F_X(x)+a \right ) f_X(x) {\rm d}x
\end{equation}
results in
\begin{equation}
\int_{a}^{b}\psi (u){\rm d}u=0. 
\end{equation}
Now, since that
\begin{equation}
\left | \frac{\psi \left ( (b-a)F_X(x)+a \right )}{\max_{\zeta \in {\rm supp} \psi} |\psi(\zeta)|} \right |\leq 1,
\end{equation}
the added perturbation is constrained to hold the inequality $0\leq \left | \varepsilon (x) \right |\leq f_X(x)$.
Thus, \textit{any} wavelet of compact support can be used to induce a {different} perturbation in the vicinity of the probability distribution initially assigned.\\
To sum up, given a random variable $X$ with CDF $F_X(x)$, a perturbation is added, which guarantees that the modified function is still a distribution arround the original CDF. This new distribution could be seen as a privatized version of the reference distribution and the privatization mechanism could be called wavelet perturbation.
\section{Moments correction due to the perturbation}
The hypothesized distribution (initial or prior distribution around which the wavelet-perturbation is introduced) has its moments defined by
\begin{equation}
\mathbb{E}(X^k)=\int_{-\infty}^{+\infty} x^k {\rm d}F_X.
\end{equation}
By introducing the perturbation defined in \ref{eq:perturbation_def}, the new (adjusted/privatized) moments are given by
\begin{equation}
\mathbb{E}_{new}(X^k):=\int_{-\infty}^{+\infty} x^k {\rm d}F_{new~X}.
\end{equation}
Thus, using
\begin{equation}
{\rm d}F_{new~X}={\rm d}F_X+\psi \left ( (b-a)F_X(x)+a \right ) {\rm d}F_X,
\end{equation}
it follows that
\begin{equation}
\mathbb{E}_{new}(X^k)=  \mathbb{E}(X^k) + \frac{1}{b-a} \int_{a}^{b} \left [ F^{-1}_X 
\left ( \dfrac{u-a}{b-a} \right ) \right ]^k \psi(u) {\rm d}u,
\end{equation}
and the second term on the right side of the previous equation accounts for a moment correction due to the introduced wavelet-perturbation.\\ Let us consider now the particular case of a perturbation in a (normalized) uniform distribution, $X \sim \mathcal{U}(0,1)$. In order to evaluate the moments of the new probability distribution $F_{new}(x)$ under the wavelet-perturbation $\psi$, with a compactly support $[0,1]$, one has
\begin{equation}
\mathbb{E}_{new}(X^k)=  \mathbb{E}(X^k) +  \int_{0}^{1}  u^k \psi(u) {\rm d}u,
\end{equation}
that is to say the moment of the wavelet \cite{Burrus} used to build the additive perturbation also {adds to} the moment of the starting distribution, because
\begin{equation}
\mathbb{E}_{new}(X^k)=  \mathbb{E}(X^k) +  \int_{-\infty}^{+\infty}  u^k \psi(u) {\rm d}u= \mathbb{E}(X^k) + M_k.
\end{equation}
In the general case, if $\psi$ has a support $[a,b] \neq [0,1]$, a modified (supported-normalized) wavelet $\psi_{[0,1]}:=\frac{\psi \left ( (b-a)x+a \right )}{b-a}$ is built and the added term is the moment of $\psi_{[0,1]}$.
\begin{equation}
\mathbb{E}_{new}(X^k)=  \mathbb{E}(X^k) +  \int_{-\infty}^{+\infty}  u^k \psi_{[0,1]}(u) du.
\end{equation}

\section{Generalizing the perturbation approach at further levels}

In the case that a (beta)-perturbation occurs over a uniform distribution [0,1], it depends on the parameters $\alpha$ and $\beta$ of the disturbing wavelet, so it is worth rewriting (via Equations~\ref{eq:perturbation}--\ref{eq:perturbation_def})
\begin{align*}
F_{new}(x)= &~~~x~~~+\Psi_{[0,1]}(x; \alpha, \beta).\\
& \text{approx + detail}
\end{align*}

This rich interpretation of wavelet theory (approximation+detail) can be generalized into the lines of a wavelet tree.

\subsection{Level-1}  Parameters: $(\alpha, \beta)$
\begin{equation}
F_{level-1}(x)=x+ \Psi_{[0,1]}(x;\alpha, \beta).
\end{equation}
Examples are such as those of Fig.~\ref{fig:fig_waveletbeta}b.
\subsection{Level-2 LH} 
Parameters: $(\alpha_L, \beta_L \vdots \alpha_H, \beta_H)$
\begin{equation}
F_{level-2}(x)=\left\{\begin{matrix}
x+ \Psi_{[0,1]}(2x;\alpha_L, \beta_L)
& 0 \leq x \leq 1/2 \\ 
x+ \Psi_{[0,1]}(2x-1;\alpha_H, \beta_H)& 1/2 \leq x \leq 1.
\end{matrix}\right. 
\end{equation}
The following is a simple example using $(\alpha_L=4, \beta_L=3 \text{ and } \alpha_H=3, \beta_H=7)$ (same parameters as in Fig.~\ref{fig:fig_waveletbeta}b, except that a set, index L, for ``the low $[0,1/2]$''  and another, index H, for ``the upper $[1/2,1]$''. But it is worth to note that different wavelets can be selected to fit different segments of the initial distribution support, for example, in a two-level perturbation, L-level can use a beta wavelet whereas the   H-level use a sombrero wavelet.
\begin{figure}[!ht]
	\centering
	\begin{subfigure}{0.4\textwidth}
		\includegraphics[scale=0.5]{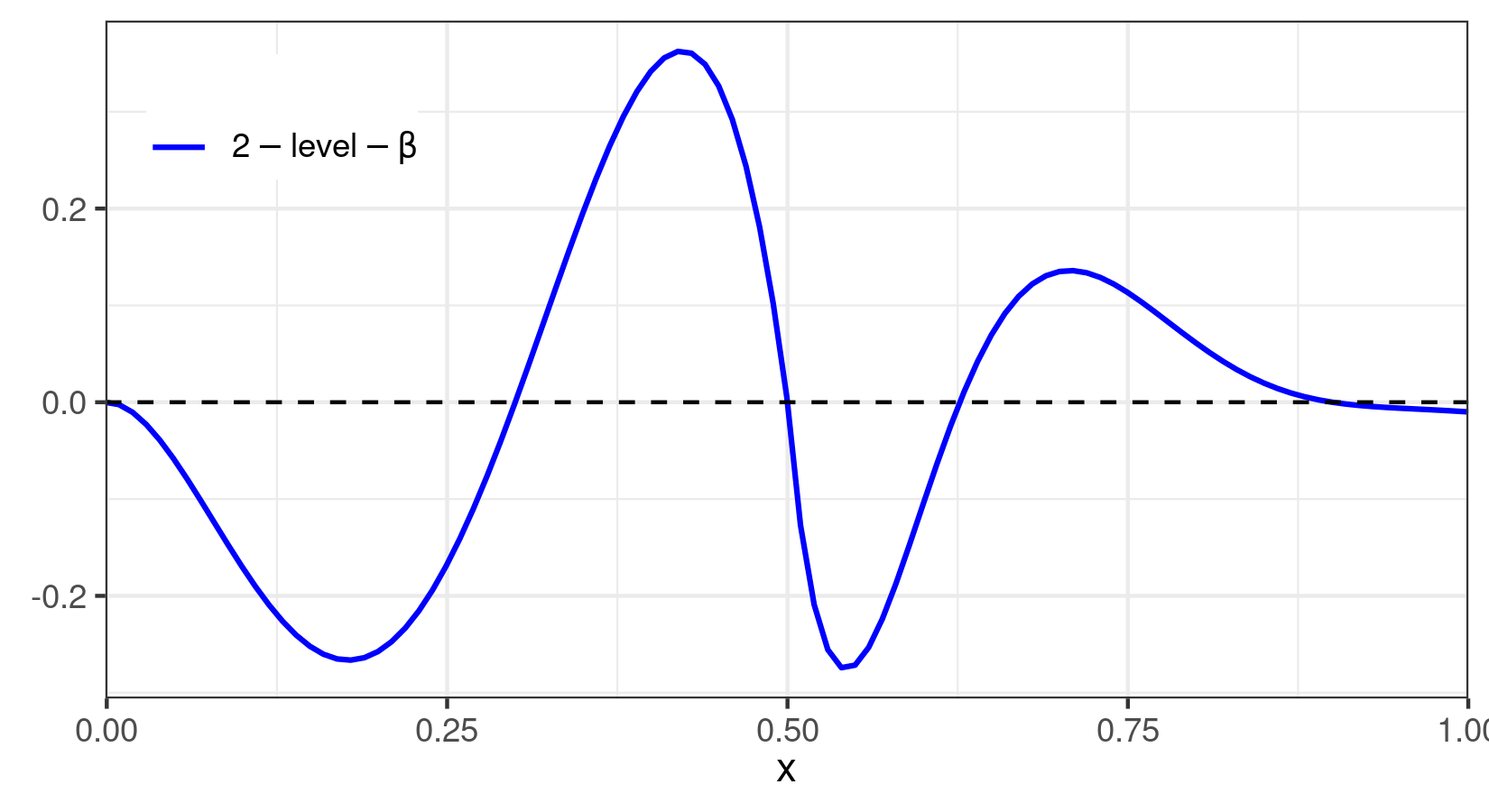}
		\captionof{figure}{2-level-beta-wavelet $\varepsilon(x)$ perturbation associated with the uniform distribution: compare to the two curves of  Fig.~\ref{fig:fig_waveletbeta}a.}
	\end{subfigure}%
	\hfill
	\begin{subfigure}{0.4\textwidth}
		\includegraphics[scale=0.5]{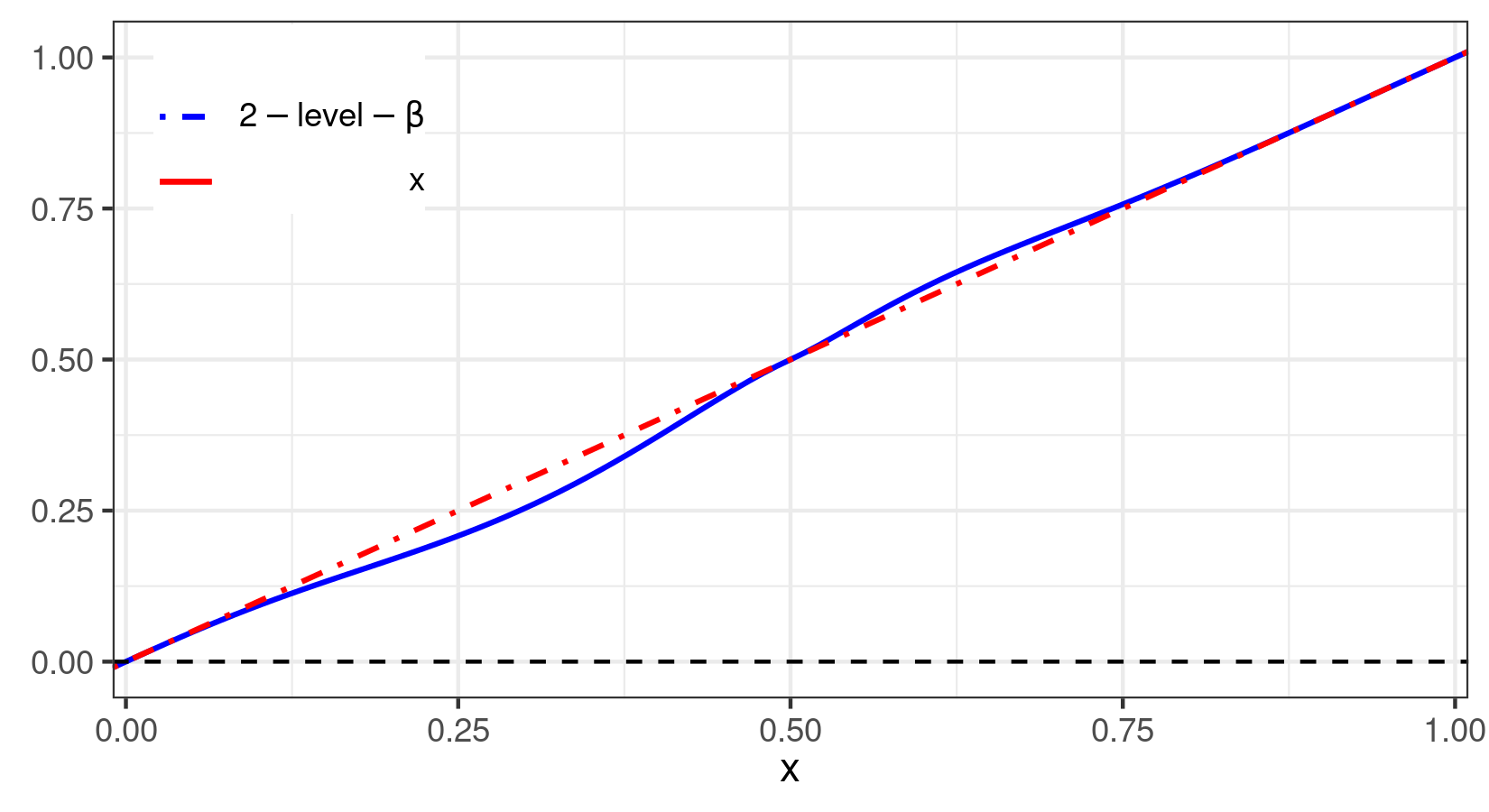}
		\captionof{figure}{2-Level-beta-perturbed distribution (\textit{line}), and uniform distribution (\textit{dashed line}).}
	\end{subfigure}
	\caption{2-Level-perturbation on a Uniform Probability Distribution Function $\mathcal{U}[0,1]$.}
	\label{fig:2level}
\end{figure}
\subsection{Level-4 LL LH HL HH}
Parameters:
\scriptsize
{$(\alpha_{LL},\beta_{LL} :\alpha_{LH},\beta_{LH} \vdots \alpha_{HL},\beta_{HL}:\alpha_{HH}, \beta_{HH})$}
\begin{equation}
F_{level-4}(x)=
\left\{\begin{matrix}
x+ \Psi_{[0,1]}(4x;\alpha_{LL}, \beta_{LL})
& 0 \leq x \leq 1/4 \\ 
x+ \Psi_{[0,1]}(4x-1;\alpha_{LH}, \beta_{HL})& 1/4 \leq x \leq 1/2\\
x+ \Psi_{[0,1]}(4x-2;\alpha_{HL}, \beta_{HL})
& 1/2 \leq x \leq 3/4 \\
x+ \Psi_{[0,1]}(4x-3;\alpha_{HH}, \beta_{HH})
& 3/4 \leq x \leq 1. \\ 
\end{matrix}\right. 
\end{equation}
\normalsize
This four-level approximation case could be  illustrated using\\
\scriptsize
{$(\alpha_{LL},\beta_{LL} :\alpha_{LH},\beta_{LH} \vdots \alpha_{HL},\beta_{HL}:\alpha_{HH}, \beta_{HH})=(4,3 \text{ : } 3, 7 ~\vdots~ 5, 3 \text{ : } 2, 7)$.}\\ 
\normalsize

\vspace{0.03 cm}
\noindent
One interpretation for this approach is to consider a distinct perturbation in each quartile of the distribution.
\scriptsize
\begin{itemize}
	\item
	1st quartile driven by $(\alpha_{LL},\beta_{LL})=(4,3)$
	\item
	2nd quartile driven by $\alpha_{LH},\beta_{LH})=(3,7)$
	\item
	3rd quartile driven by $(\alpha_{HL},\beta_{HL})=(5,3)$
	\item
	4th quartile driven by $(\alpha_{HH},\beta_{HH})=(2,7)$.
\end{itemize}
\normalsize
\section{Concluding Remarks}
This paper provide a new method to build an additive wavelet-based perturbation (privacy mechanism) to modify a given continuous probability distribution. The initial guess can then be perturbed as some sort of ``prospecting within the ensemble of possible probability distributions around the starting distribution''. A procedure is also offered to fit four different perturbations, one in each quartile of the distribution, which can be quite attractive in data fittings. 
%
\section{Funding}
This research was partially supported by the National Council
for Scientific and Technological Development (CNPq) through
the grant 305305/2019-0 (RO), and Comissão de Aperfeiçoa-
mento de Pessoal do Nível Superior (CAPES), from the Brazilian government; and by FONDECYT, grant number 1200525
(V. Leiva), from the National Agency for Research and Development (ANID) of the Chilean government under the Ministry
of Science, Technology, Knowledge, and Innovation


\end{document}